\documentclass[aps,pra,showpacs,amsart,twocolumn,10pt]{revtex4-2}

\usepackage{graphicx}
\usepackage{amsmath,amssymb}
\usepackage{color}
\usepackage[colorlinks=true,linkcolor=blue,urlcolor=blue,citecolor=blue]{hyperref}
\usepackage{mathrsfs}
\usepackage[utf8]{inputenc}
\usepackage{bm}
\usepackage{physics}
\usepackage{mathtools}
\usepackage{mathptmx}
\usepackage{helvet}

\usepackage{hyperref}

\makeatletter
\newcommand{\printfnsymbol}[1]{%
  \textsuperscript{\@fnsymbol{#1}}%
}
\makeatother

\begin{document}

\title{Site-Resolved Imaging of Bosonic Mott Insulator of $^7$Li atoms}
\author{Kiryang Kwon}
\author{Kyungtae Kim}
\author{Junhyeok Hur}
\author{SeungJung Huh}
\author{Jae-yoon Choi}
\email[Corresponding author: ]{jaeyoon.choi@kaist.ac.kr}
\affiliation{Department of Physics, Korea Advanced Institute of Science and Technology, Daejeon 34141, Korea \\}
\date{\today}

\begin{abstract}
We demonstrate a single-site and single-atom-resolved fluorescence imaging of a bosonic Mott insulator of $^7$Li atoms in an optical lattice.
The fluorescence images are obtained by implementing Raman sideband cooling on a deep two-dimensional square lattice, where we collect scattered photons with a high numerical aperture objective lens. 
The square lattice is created by a folded retro-reflected beam configuration that can reach 2.5~mK lattice depth from a single laser source. 
The lattice beam is elliptically focused to have a large area with deep potential.
On average 4,000 photons are collected per atom during 1~s of the Raman sideband cooling, and the imaging fidelity is over 95$\%$ in the central 80$\times$80 lattice sites. 
As a first step to study correlated quantum phases, we present the site-resolved imaging of a Mott insulator.
Tuning the magnetic field near the Feshbach resonance, the scattering length can be increased to 680$a_B$, and we are able to produce a large-sized unity filling Mott insulator with 2,000 atoms at low temperature.
Our work provides a stepping stone to further in-depth investigations of intriguing quantum many-body phases in optical lattices.

\end{abstract}


\maketitle



\section{Introduction}

Ultracold atoms in optical lattices have been an outstanding platform for quantum simulations to study complex many-body quantum problems in a wide range of research~\cite{Gross2017,Schafer2020}. 
With the advent of the high-resolution imaging systems, the so-called quantum gas microscopes, the individual atoms that constitute the ``synthetic" quantum materials can be imaged with near unity fidelity~\cite{Bakr2009,Sherson2010}.
This microscope system not only reveals intriguing features of many-body quantum states by directly measuring correlations between particles~\cite{Endres2011}, but also control the atomic quantum state in a minimal physical unit~\cite{Weitenberg2011}, and thereby deepens our understandings of complex quantum phenomena.
The site-resolved imaging technique been demonstrated for bosonic, $^{87}$Rb~\cite{Bakr2009,Sherson2010} and $^{174}$Yb~\cite{Miranda2015,Yamamoto2016}, and fermionic,$^6$Li~\cite{Parsons2015,Omran2015} and $^{40}$K~\cite{Cheuk2015,Haller2015,Edge2016}, atoms in a two-dimensional square lattice. 
Recently the quantum gas microscope has been realized in a triangular lattice for bosonic $^{87}$Rb~\cite{Yamamoto2020} and fermionic $^6$Li~\cite{Yang2021} atoms, facilitating the quantum simulation of frustrated systems~\cite{Wannier1950,Anderson1987}.

Among various microscope systems, however, only the Fermi gas microscopes ($^6$Li and $^{40}$K) have used magnetic Feshbach resonance, which can change  the $s$-wave scattering length and the interactions in the system~\cite{Chin2010}. 
Taking advantages of the atomic properties, the Fermi gas microscopes have investigated a repulsive~\cite{Parsons2016,Boll2016,Cheuk2016,Brown2017} and an attractive~\cite{Mitra2018} side of the Fermi-Hubbard model, observing a long-range anti-ferromagnetic spin orders at half filling~\cite{Mazurenko2017} and evidence of a pseudo gap~\cite{Brown2020}.
Like the Fermi gases, the bosonic $^7$Li isotopes display a broad Feshbach resonance near 738~G~\cite{Gross2010,Dyke2013} with a wide interaction tunability~\cite{Pollack2009a}. 
Additionally, the light atom mass $m$ provides the advantages of large recoil energy, so that its tunneling time can be faster than the other bosonic species with a broad resonance, i.e.,  $^{39}$K($^{133}$Cs) by a factor of 5.6(19).
Because of the advantageous atomic property, growing interest has recently focused on $^7$Li atoms in optical lattices~\cite{Geiger2018,Fujiwara2019,Amato2019,Jepsen2020,Jepsen2021a,Jepsen2021b}.
Position-space Bloch oscillations have been directly observed in a tilted optical lattices~\cite{Geiger2018}, and the long-range transport of Bose condensates can be realized with additional periodic modulation of the lattice amplitude~\cite{Fujiwara2019}. 
The system also provides interaction tunability between two atoms in a different sub-hyperfine levels~\cite{Amato2019}, introducing a versatile experimental platform to study spin-1/2 Heisenberg XXZ spin models~\cite{Jepsen2020}. 
The spin interaction anisotropy dramatically changes the spin transport dynamics~\cite{Jepsen2020,Jepsen2021a}, and a long-lived helix state is recently observed under a specific interaction anisotropy~\cite{Jepsen2021b}.

\begin{figure}
\centering
\includegraphics[width=1\linewidth]{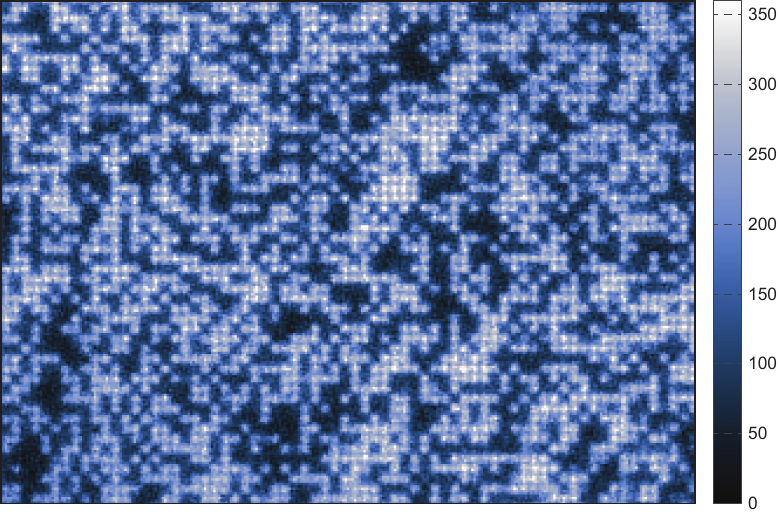} 
\caption{Fluorescence imaging of $^7$Li atoms in a two-dimensional square lattice. About 4,000 photons per atom are collected during 1~s of the Raman imaging process. The color bar represents the photon count in a low-noise camera. The field view of the image is $52.7\times37.4~\mu$m$^2$ ($70\times50$ sites). }
\label{Ffluo}
\end{figure}

In this work, we report the quantum gas microscopy of $^7$Li atoms in a two-dimensional (2D) square optical lattice (Fig.~\ref{Ffluo}). 
We have achieved single-site resolved fluorescence imaging by employing a Raman sideband cooling in a deep two-dimensional lattice. 
A single high-power laser at 1,064~nm wavelength is used to create the square optical lattice, where we exploit a folded retro-reflected beam configuration~\cite{Sebby2006,Brown2017}. 
At its maximal beam power, the lattice depth is 2.5~mK and the sideband frequency is 1.5~MHz, which satisfy the Lamb-Dicke condition.
Additionally, a tightly focused light sheet potential confines the atoms in the image plane of the objective lens during the imaging.
About 4,000 photons are collected per atom during the 1~s of imaging time, and we have achieved high imaging fidelity ($>95\%$) over a large area of $80\times80$ sites. 
To enter into a strongly interacting regime, we add a vertical accordion optical lattice that can dynamically change its lattice spacing from $10~\mu{}$m to $2~\mu{}$m. 
We select a single plane of the vertical lattices using the light sheet and observe a Mott insulating phase after ramping up the square lattice.
Using the Feshbach resonance, we increase the $s$-wave scattering length to 680$a_B$ and produced a large sized unity filling Mott insulator with 2,000 atoms and low entropy per particle, $S/(Nk_B)=0.31(1)$.

This paper is organized as follows. 
In Sec.\ref{setup}, we present the experimental setup of the quantum gas microscope including lattices, Raman sideband cooling, and high-resolution imaging system. 
The experimental sequences to prepare a single layer of 2D degenerate quantum gases in three dimensions are also introduced. 
In Sec.\ref{Imaging}, fluorescence images and site occupation analysis algorithm are described. 
In Sec.\ref{MI}, we report the single-site resolved imaging of the Mott insulator with tunable interaction by using the Feshbach resonance, and conclude with an outlook in Sec.\ref{Conclusion}.

\section{\label{setup}Experimental Setup}
\subsection{Optical Transport}

The experiment begins by transporting pre-cooled $^7$Li gases captured in a vacuum chamber (main chamber) to an adjacent vacuum chamber (science chamber), where a high-resolution imaging system is installed.
The central displacement between the two vacuum chambers is 460~mm, and we transport the cold thermal gases in the main chamber to the science chamber by moving the focal position of an optical potential (transport trap)~\cite{Gustavson2001,Gross2016}. 
The transport trap is made of a 1,064~nm laser beam and has a pancake-shaped geometry with a vertical (lateral) beam waist of $19(535)~\mu${}m at the main chamber. 
By mounting a cylindrical lens on an air-bearing stage (ABL1000, AeroTech), we are able to translate the focal position upto 1~m. 
The vertical beam waist at the science chamber is increased to 58$~\mu${}m because of beam clipping at the entrance of the viewport.
To maintain a similar trap depth during transport, we weakly focused the lateral beam waist to 300$~\mu${}m at the target distance. 

In the experiment, the cold gases are prepared after an evaporation cooling in a magnetic trap~\cite{Kim2019}.
Ramping up the transport beam to 3.3~W in 200~ms at the end of the cooling process, we are able to load $N\simeq 1.0\times 10^7$ of thermal gases at $T=2.1(3)~\mu{}$K.
To reduce heating and atom loss, we flip the spin state from an upper hyperfine state ($|F=2,m_F=2\rangle$) to a lower hyperfine state ($|F=1,m_F=1\rangle$) in the optical potential.
Then, we increase the beam power to 12.3~W and move its focal point to follow a sigmoidal curve that accelerates with a maximum of 2.1~$m/s^2$ to the midpoint and symmetrically decelerates afterwards.
The acceleration can reduce the trap depth along the optical axis, and the minimal trap depth during the transport is  $\sim 20~\mu$K. 
After 1.3~s of transport time, $N\simeq 3.0\times 10^6$ cold atoms can be prepared in the science chamber.

\begin{figure*}
\centering
\includegraphics[width=1\linewidth]{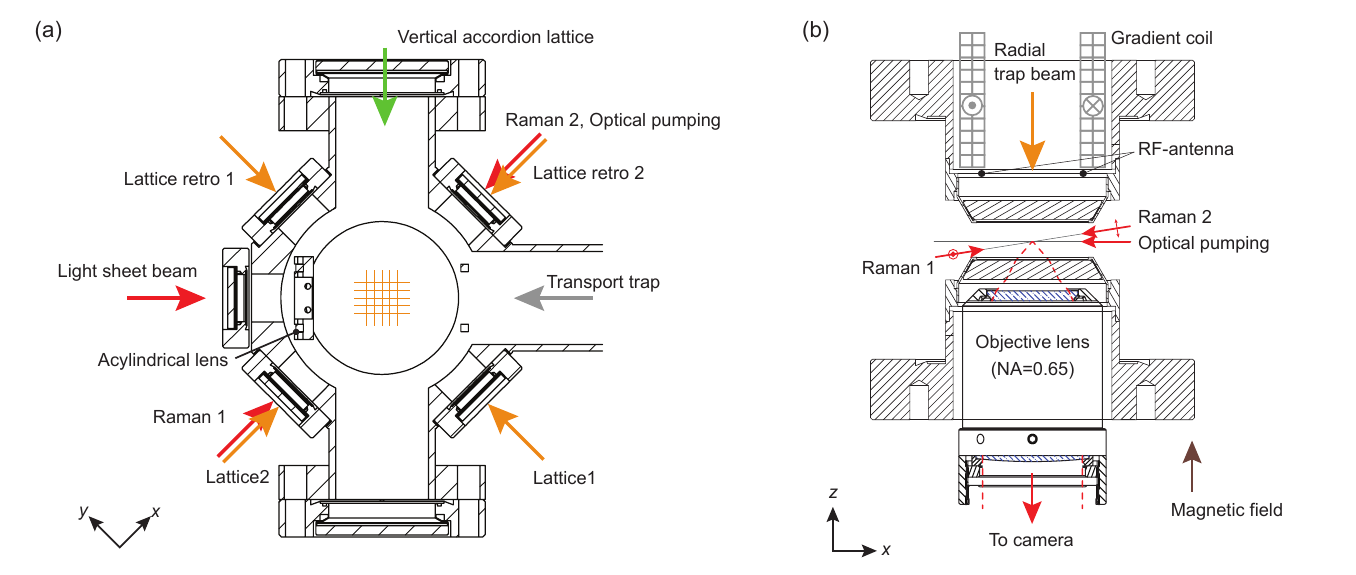} 
\caption{Experimental setup for the quantum gas microscopy of $^7$Li atoms. 
(a) A cross-section view of the stainless steel vacuum chamber in the $x$-$y$ plane and a sketch of various beam arrangements.
A two-dimensional square lattice (orange grid) is created by a single laser beam at 1,064~nm wavelength, where a incoming light (lattice1) is retro-reflected with a $90^{\circ}$ folding angle. 
An acylindrical lens is mounted inside the vacuum chamber because of its short working distance (23.6~mm). 
The Raman cooling beams and optical pumping light are sent through the view port for one of the lattice beams, having a projection angle of $45^{\circ}$ with respect to the lattice axis.
(b) A schematic sketch of the apparatus in a side view. 
Recessed bucket-shaped windows are installed for high resolution imaging.
 The objective lens is designed to correct the aberration from 5 mm thick fused silica window.
 A high numerical aperture (NA=0.65) objective lens is placed on a high precision objective scanner (not shown), which is secured on the tilt stage. 
Raman beams (red arrows) propagate with an angle $\theta_R=9^{\circ}$ out of the horizontal plane and are both linearly polarized. 
A weak magnetic field ($< 50$~mG) is applied along the vertical axis. 
An optical trap beam for the radial confinement is provided from the top viewport (orange arrow).  
A 22-turn water-cooled coil is mounted on the top bucket window and produces a field gradient along the vertical axis.
A single-turn radio frequency wire is attached to the gradient coil. 
The Rabi frequency for magnetic sub-level transition in the $|F=1\rangle$ state is about 3~kHz at 10~W of rf-power.
}
\label{Fsetup}
\end{figure*}

\subsection{Quantum gases in a single layer of the optical lattice} 

To prepare 2D degenerate Bose gases, we transfer the thermal clouds in the transport trap into a tightly focused light sheet beam. 
Inserting an acylindrical lens (Asphericon) inside the vacuum chamber [Fig.~\ref{Fsetup}(a)], a 1,064~nm laser beam is focused to have $1/e^2$ beam waists \textcolor{black}{with} $4.4(87)~\mu$m in the vertical (lateral) direction. 
An evaporation process is carried out at a constant scattering length $a_s=138(6)a_B$, where $a_B$ is the Bohr radius.
A field gradient of 37~G/cm is applied along the vertical axis to reduce the potential depth without losing its harmonic confinement [Fig.~\ref{Fsetup}(b)], and thereby produce efficient evaporation cooling in the optical trap~\cite{Hung2008}.  
After 300~ms evaporation, pure condensates containing 10,000 atoms can be produced.

Then, we load the atoms into a single layer of the vertical lattice. 
The vertical lattice is made by a 532~nm laser beam, and we employ an accordion lattice setup to ensure a stable loading process~\cite{Mitra2016,Ville2017}.
The light source is obtained by a single-pass frequency doubling of 1,064~nm light with a nonlinear crystal (PP-Mg:SLT, Oxide). 
About 1~W of green beam is delivered to the science chamber using a high power air gap optical fiber (QPMJ-A3AHPM, OZ optics).
Splitting the output beam into two parallel propagating beams and focusing them with an aspheric lens (AL50100-A, Thorlabs), we create the vertical lattice at the crossing position. 
To make a large Mott insulator, the two beams are elliptically focused with a beam waist of 34(300)~$\mu${}m along the vertical(lateral) direction. 
The lattice spacing can be tuned from 12~{$\mu$}m to 2~{$\mu$}m by changing the relative distance between the two beams (XMS50-S, Newport)~\cite{Ville2017}. 
Starting from a large lattice spacing, we are able to load the quantum gases into a single node of the vertical lattice, and then adiabatically compress the atoms.
The vertical frequency after compression is 23(2)~kHz. 
During the loading process, we gradually turned off the light sheet potential and confined the atoms in the two dimensional plane using an additional radial trap beam [Fig.~\ref{Fsetup}(b)].
The final number and temperature of the atoms can be controlled by reducing the trap depth of the radial trap beam.

\subsection{Two-dimensional square optical lattice}

Fluorescence imaging of atoms in optical lattices requires a high power laser system to preserve the positional information of the atoms during the light scattering process.
Moreover, light atoms like $^6$Li and $^7$Li are vulnerable to laser noise because of its high on-site frequency~\cite{Savard1997,Gehm1998}, and thus a high-power and low-noise laser system has to be implemented for the quantum gas microscope~\cite{Blatt2015}. 
We obtain the high power 1,064 nm lattice beam from a Yb-doped fiber amplifier system (ALS-IR-1064-50-CC, Azurlight), seeded by a narrow linewidth laser (Mephisto2000 NE, Coherent).  
The maximal beam power is 45~W after an internal isolator (30~dB), and a 30~dB high power isolator (Pavos-Ultra, EOT) is added to prevent any feedback from a reflected light.

\begin{figure*}
\centering
\includegraphics[width=1\linewidth]{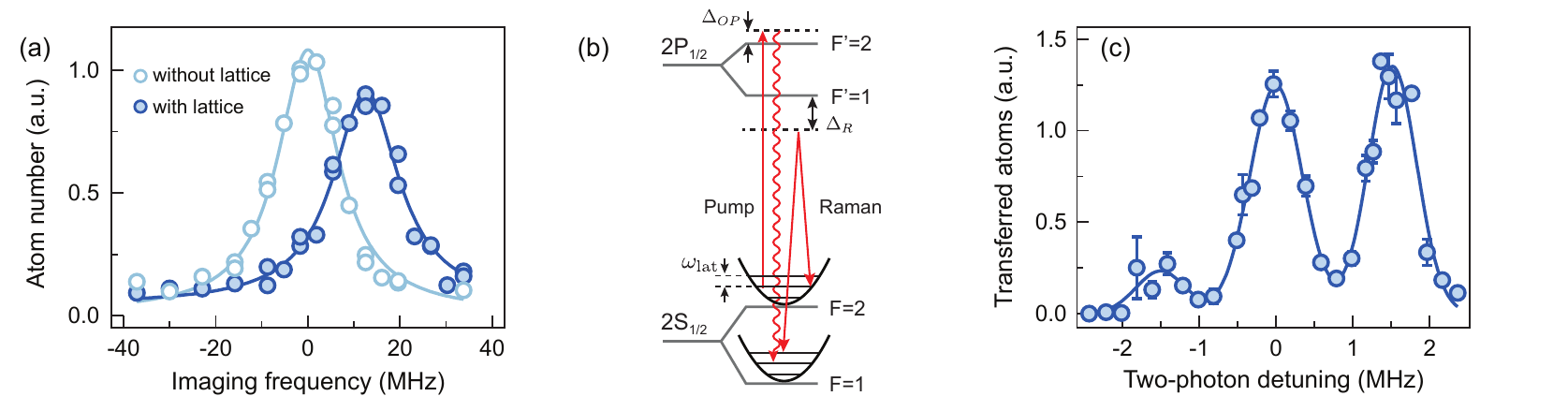} 
\caption{(a) Estimation of the 2D lattice depth. 
Absorption spectra of thermal atoms in the $F=1$ hyperfine level with (solid dark blue) and without (open light blue) the lattice potential.
Solid lines are the Lorentzian fit to data, where we extract the 12.6~MHz light shift by the optical lattice.
(b) Schematic diagram of the Raman sideband cooling.  
Two-photon Raman transition transforms the hyperfine spin state of trapped atoms from $|F=1\rangle$ to $|F=2\rangle$ state while simultaneously removing one vibrational quantum number. 
Shining a pumping beam, atoms are optically pumped back into the lower hyperfine spin state, and we collect scattered photons and image the site occupation. 
The Raman beams are red-detuned from the $D_1$ transition by $\Delta_R=7$~GHz and the optical pumping beam is blue-detuned from the $|F'=2\rangle$ state by $\Delta_{OP}=3$~MHz in a free space. 
(c) Two-photon Raman transition under a deep square lattice potential ($4000E_{\rm r}$). 
The central peak corresponds to the carrier transition with the same vibrational level, while the sidebands represent interband transition. 
From the sideband peak positions, we determine the on-site frequency to be $\omega_{\rm lat}=2\pi\times1.5(1)$~MHz. 
The red sideband is used for Raman cooling and imaging. 
The data is taken with 3 independent experimental runs, and the error bars represents 1 standard deviation of the mean.
}
\label{Framancool}
\end{figure*}

The optical lattice is created by a folded retro-reflected standing wave configuration, where four vertically polarized beams are interfered to make a square lattice with its spacing $a_{\rm{lat}}=1,064/\sqrt{2}=752~{\rm nm}$. 
This provides an economical solution to realize deep optical lattices with limited optical power~\cite{Sebby2006}, which was first implemented in the $^6$Li quantum gas microscope ~\cite{Brown2017} and recently extended to a triangular lattice geometry~\cite{Yang2021}.
In our setup, we use an elliptically focused ($138~\rm{\mu m}\times40~\rm{\mu m}$) lattice beam to increase the beam size in the $x$-$y$ plane with sufficient potential depth for the fluorescence imaging. 
The folding angle between the two lattice arms is nearly 90$^\circ$, and we optimize the lattice overlap between the incoming and retro-reflecting beam by studying the Kapitza-Dirac diffraction of a Bose-Einstein condensate.
At a given lattice power, the number of atoms in the zero-th order is minimized after a short pulse of 2$~\mu{}$s.  
The lattice depth at its full power (30~W) is calibrated by measuring the light shift from the lattice beam. 
Studying a resonant $D_1$ transition line ($\ket{F=1}\rightarrow \ket{F'=2}$), we observe a 12.6~MHz of blue-shift under the lattice beam [Fig.~\ref{Framancool}(a)]. 
Considering the ac polarizabilities of the $2S_{1/2}$ and $2P_{1/2}$ states at 1,064~nm~\cite{Safronova2012}, it corresponds to the $3,950E_{\rm r}$ for the ground state, and the on-site frequency of the lattice is $\omega_{\rm lat}/2\pi=1.5$~MHz.

While most of the tunneling dynamics occurs at shallow lattice depth (tens of $E_{\rm r}$), the atoms have to be fully pinned in each site at thousands of $E_{\rm r}$ lattice depth for fluorescence imaging. 
Here, the $E_{\rm r}=h^2/8ma_{\rm lat}=h\times12.6$~kHz is the recoil energy of the lattice, and $h$ is Planck's constant.
To control the lattice depth over such a wide dynamic range, we employ a two-stage power control system: (1) regulating the rf power input to the acousto-optical modulator (AOM-3080-199, Gooch and Housego). (2) variable power attenuation from its full power by rotating a $\lambda/2$-waveplate (DDR25, Thorlabs). 
Combining two stages, we have over 30dB dynamic range. 
Operating the system at low power (100~mW), the light intensity is stablized by an analog PID controller (SIM960, SRS), and the relative intensity-noise of the lattice beam is below $-125$~dBc/Hz in the $10-100$~kHz frequency range.
Before taking the fluorescence imaging, the lattice depth is first increased to $100E_{\rm r}$ in $300~\mu$s, and then we rotate the waveplate to have $4,000E_{\rm r}$ potential depth in 100~ms.

\subsection{Raman sideband cooling}

To maintain the atomic position in each lattice site during the fluorescence imaging, we apply Raman sideband cooling in the deep optical lattice. 
The scheme is based on a two-photon Raman sideband transition between two lowest hyperfine ground states while lowering the vibrational state in each lattice potential.
The cooling cycle is completed by optically pumping the atomic hyperfine state to its initial state without changing the motional state.
After many cycles, most of the atoms can be populated in the motional ground state, which is the dark state to the Raman laser light.
Heating can occur when the vibrational state changes during the repumping process so that it requires the Lamb-Dicke regime and a tight confinement of the lattice ($\omega_{\rm lat}\sim 2\pi\times1$~MHz for Lithium atoms).

The experimental scheme for the Raman sideband cooling of $^7$Li atoms in shown in Fig.~\ref{Framancool}(b).  
The atoms in the lowest hyperfine state $|{F=1}\rangle$ are transferred to the upper hyperfine state $|{F=2}\rangle$ via the two-photon transition.
In the experiment, we apply a linearly polarized laser light, which is 7~GHz red tuned to the $D_1$ transition line, and is focused with a beam waist of $100~\mu{}$m at the lattice center. 
The incoming beam is retro-reflected [Fig.~\ref{Fsetup}(a)] after passing through an acousto-optic modulator (MT350-B120, AA Opto Electronics) to control Raman detuning.
The polarization of the retro-beam is also linear, but is orthogonal to the incoming beam.
The optical pumping beam is circularly polarized and is applied through the diagonal axis of the lattice. 
It is 3~MHz blue-detuned to the $|{2S_{1/2} F=2}\rangle\rightarrow|{2P_{1/2} F'=2}\rangle$ transition in a free space.
The magnetic field is compensated to below 50~mG, and we set its direction to point along the $z$ axis.

From the sideband spectroscopy, the level spacing of the deep lattice is determined to $\omega_{\rm lat}= 2\pi\times1.5$~MHz [Fig.~\ref{Framancool}(c)], and the Lamb-Dicke parameter for the Raman transition is $\eta_{R}\simeq  k_R\sqrt{\hbar/(2m\omega_{\rm lat})}=0.29$, where $k_R$ is the Raman wave vector projected along the lattice axis. 
The Lamb-Dicke parameter for the pumping beam is $\eta_{\rm OP}\simeq 0.2$.
The Raman cooling in the vertical axis is inefficient, as the maximal confinement of the sheet potential is $\omega_{\rm sh}\simeq 2\pi\times100$~kHz. 
Still, this configuration has provided sufficient imaging fidelity~\cite{Brown2017,Yang2021}.
This is because the vibrational state along the $z$ axis can be coupled to the motional state in the square lattice by the small tilting angle of the Raman beam, having effective cooling in the vertical axis.

\subsection{High-resolution Imaging system}

A high numerical aperture (NA=0.65) objective lens (54-26-20@532-1,064, Navitar), which collects scattered photons during the optical pumping process, is placed below a bucket-shaped window~[Fig.~\ref{Fsetup}(b)].
The objective lens is designed to compensate wavefront distortion from the vacuum glass (fused silica, 5~mm thickness) when the lens axis is normal to the surface of our vacuum window and to have diffraction limited performance at 671-1,064~nm with field of view 150$\times$150$\mu$m$^2$.  
The working distance of the objective is 13~mm, and its effective focal length is 20~mm. 
The high-resolution imaging system is completed by placing an achromatic doublet (AC508-750-B, Thorlabs) with a focal length of 750 mm, and the photon number is counted by a low-noise CMOS camera (Zyla 4.2 PLUS, Andor) with a magnification factor of 37.5. 
To adjust the objective plane, we use a high precision objective piezo scanner (P-725.1CA, Physik Instrumente) and mount them on a multi-axis tilt platform (M-37, Newport).
We note that the flat surface of the objective lens helps the initial alignment procedure [Fig.~\ref{Fsetup}(b)] with the aid of wavefront interferometry.  
An incoming laser light towards the objective can be reflected on the flat lens surface, interfering with the reflected beam on the bottom window. 
Making co-centric interference ring patterns and reducing its fringe number, we are able to align the axis of the objective lens normal to the vacuum window within 1 mrad uncertainty.


\section{\label{Imaging}Single site-resolved Imaging}
\subsection{Imaging resolution}

\begin{figure}
\centering
\includegraphics[width=1\linewidth]{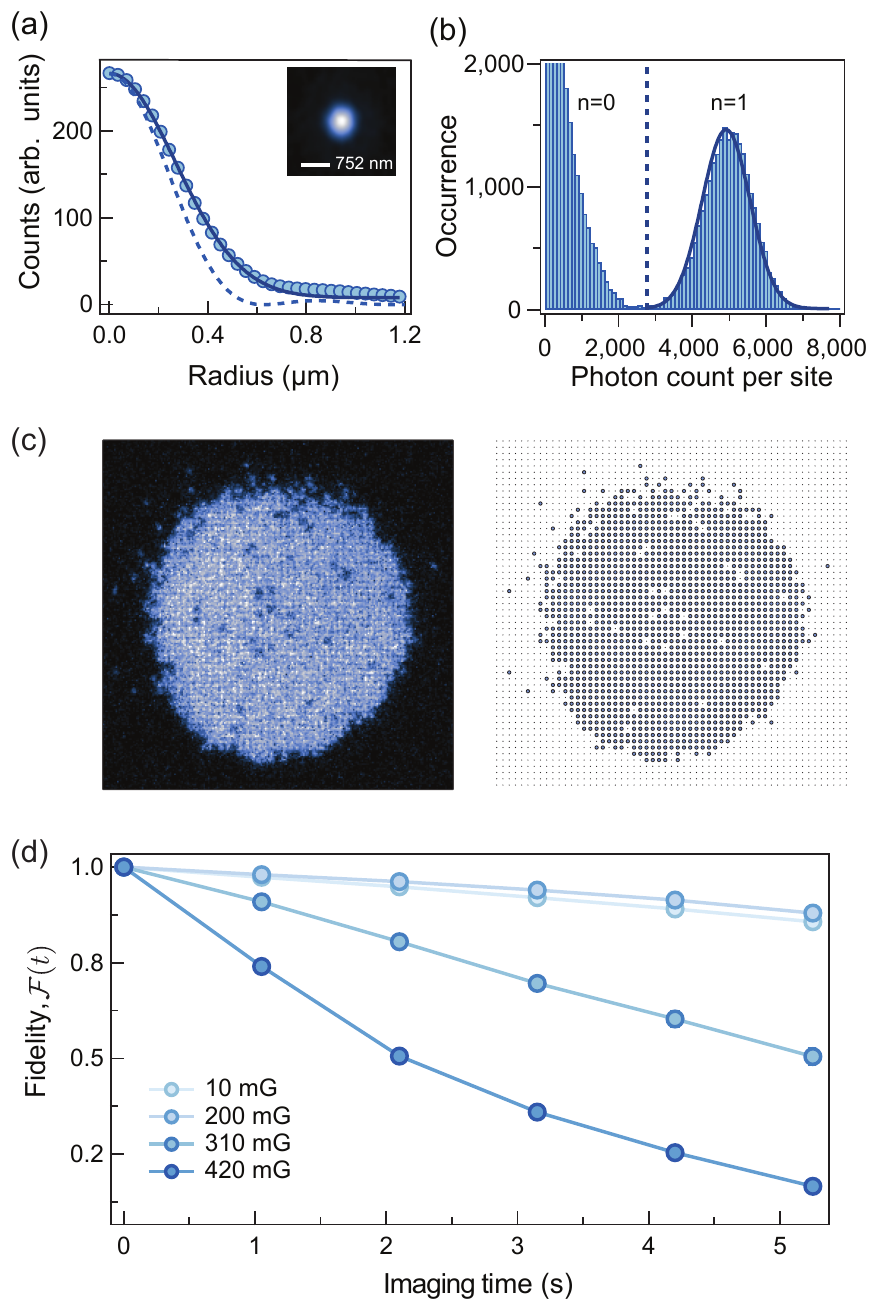} 
\caption{Image analysis. (a) Azimuthal average profile of the measured point spread function (blue solid circle). 
The solid line is the Gaussian fit to the data, and the dashed line is the radial profile of a diffraction-limited Airy function for our objective lens $\rm{NA}=0.65$.
Inset shows averages of isolated single atoms using subpixel shifting~\cite{Bakr2009}. 
(b) The histogram of photon counts from individual atoms in each lattice sites. 
The two peaks are well-separated, representing $n=0$ and $n=1$ occupation in the lattice. 
The distribution for occupied sites follows a normal distribution (solid line) with its mean $\mu=4,900$ and standard deviation $\sigma=91	0$.
The dashed line represents the threshold photon count for determining a site occupation.
(c) Using a reconstruction algorithm, described in the main text, a fluorescence image of a unity filling Mott insulator (left) can be converted to a binary occupation map (right) with high precision ($99.5\%$). 
Each filled circle represents a single atom, and the lattice sites are marked by dots.
(d) The effect of an external magnetic field during the Raman sideband cooling and imaging process. The imaging fidelity shows a similar performance when the external bias field is $B<200$~mG. 
}
\label{FigRecon}
\end{figure}

Figure~\ref{Ffluo} displays the fluorescence imaging obtained by continuously operating the Raman sideband cooling in the deep lattice. 
During the imaging process, we freeze the atoms in the image plane of the objective lens by increasing the depth of the sheet potential to 0.83~mK.
Even without post image processing, we are able to identify the square lattice structure.
To characterize and to optimize the detection fidelity, we have measured the point spread function (PSF) of our imaging system. 
Preparing sparsely filled cold atoms in the two-dimensional lattice, isolated images of single atoms can be obtained [Fig.~\ref{FigRecon}(a)].
The PSF is well approximated by a 2D Gaussian function and its full width half maximum (FWHM) is 630~nm.
This corresponds to 19\% larger than the diffraction limit, but still smaller than the lattice spacing ($a_{\rm lat}=752$~nm), so that we have achieved single-site resolution.

\subsection{Imaging fidelity analysis}

To convert the fluorescence images to a 2D array of atom occupation, we apply the following reconstruction algorithm.
Taking Fast Fourier Transform (FFT) of the raw images, we first find the lattice vectors and relative phases between them.
The angle between the lattice vectors is $90.1(1)^\circ$, close to an ideal square lattice.
The phase information of the lattice is extracted by fitting a projected image along one of the principal axes to a sinusoidal function with a Gaussian envelope~\cite{Yamamoto2020,Yang2021}.
After constructing the lattice structure, we partition the raw images site by site and deconvolute the images using the measured PSF.
Summing up the photon counts in each lattice site $N_{\rm site}$, we plot its histogram [e.g., Fig.~\ref{FigRecon}(b)]. 
It shows two peaks near $N_{\rm site}=0$ and $N_{\rm site}\simeq 5,000$, representing an empty and occupied site, respectively.
Setting the threshold count for number occupation by the intersection point between the peaks, we accurately extract the site occupation of the fluorescence images [Fig.~\ref{FigRecon}(c)].
From the analysis, we conclude that on average 4,000 photons per atom are collected during the 1~s of exposure time.
It corresponds to a fluorescence scattering rate is 50~kHz given that the collection efficiency of our imaging system is $\eta_{\rm img}\simeq 0.08$.

\begin{figure*}
\centering
\includegraphics[width=0.9\linewidth]{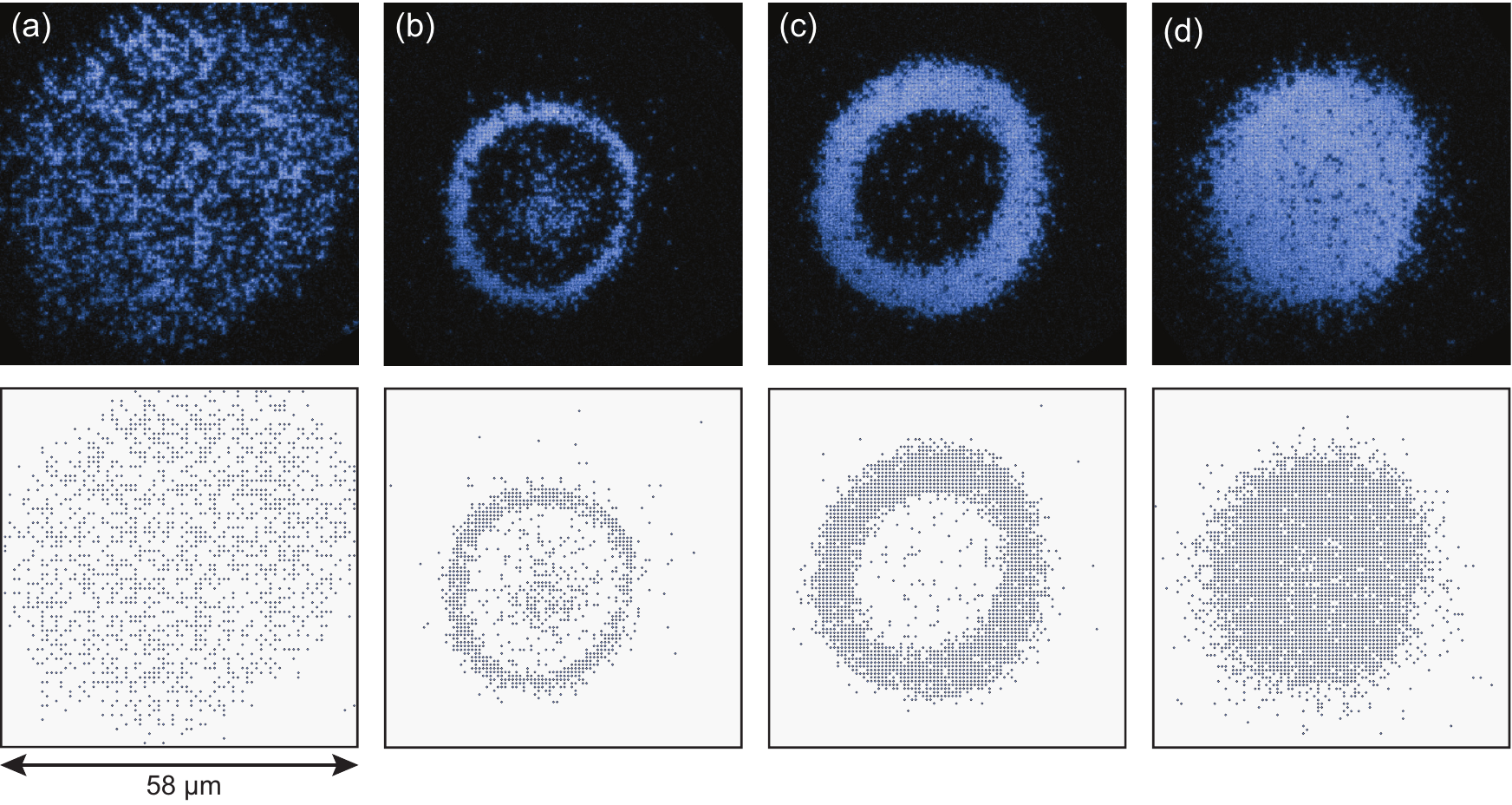}
\caption{Single-site resolved  images of superfluid and Mott insulating phases. 
The first row represents raw fluorescence images obtained by the Raman sideband cooling, and the second row displays the parity-projected atom number extracted from the reconstruction algorithm. 
(a) The BEC confined in a harmonic potential, and (b)-(d) Mott insulators under lattice potential at 24$E_{\rm r}$.
The central filling $n$ of the Mott insulator can be tuned by changing the scattering length $a_{s}$, such that (b) $n=3$ for $a_{s}=130a_B$, (c) $n=2$ for $a_{s}=280a_B$, (d) $n=1$ for $a_{s}=680a_B$. 
The temperature of the unity filling Mott insulator is $T=0.1U/k_B$.
}
\label{FigMI}
\end{figure*}

The imaging fidelity is investigated by studying the hopping and loss rate of the atoms during the imaging process.
After preparing a uniform array of atoms or unity filling Mott insulator, we take 6 successive images with 1~s of exposure time and 50~ms of interval time. 
The imaging fidelity is defined as 
\begin{equation}
\mathcal{F}(t)=\frac{\sum_{(i,j)}n_{ij}(0)n_{ij}(t)}{\sum_{(i,j)}n_{ij}(0)},
\end{equation}
where $n_{ij}(t)$ is the atom occupation number at $(i,j)$ lattice position at time $t$.

With the image analysis protocol we optimize the parameters for the Raman sideband cooling process in a way that maximizes the imaging fidelity with sufficient fluorescent count to detect the site occupation.
We observe that the fluorescent photon count increases with stronger Raman beam intensities and smaller detuning, but it also promotes off-resonant photon scattering processes, and reduces the imaging fidelity. 
At the optimal conditions, the Raman beams have 1~mW and 0.6~mW of beam power with $\Omega_R=2\pi\times$ 50~kHz of two-photon Rabi frequency.
Moreover, we apply a sinusoidal modulation of the sideband frequency so that the two-photon detuning of the Raman beam follows, $\delta=\nu_{\rm hfs}-\omega_{\rm lat}/2\pi+A_{\rm mod}\sin(2\pi\nu_{\rm mod}t)$. 
Here, $\nu_{\rm hfs}=803.5$~MHz is the hyperfine splitting energy of the ground state, $A_{\rm mod}=800$~kHz and $\nu_{\rm mod}=500$~kHz are the amplitude and frequency of the modulation, respectively.
The frequency modulation can counteract the reduction in sideband frequency caused by the Gaussian beam profiles of the lattice beam, and thus provides a large imaging area.
Under these conditions, we have an imaging fidelity of 96.1(3)\% with a loss rate of 3.2(4)\% and hopping rate of 0.7(4)\% [Fig.~\ref{FigRecon}(d)].


\section{\label{MI} Site-resolved imaging of Mott Insulator}

Equipped with the site-resolved imaging system, we are able to directly observe the superfluid and Mott insulator phases in the two-dimensional square optical lattice (Fig.~\ref{FigMI}). 
Preparing a few thousands atoms in the single plane, we ramp up the 2D lattice potential in 50~ms. 
At low lattice depth (1$E_{\rm r}$), the particle occupation fluctuates from each independent experimental run, and the variance of the parity-projected atom number is saturated $\sigma_{i,P}^2\simeq 0.24$ at the trap center. 
These are the characteristic features of a BEC in a lattice, where the mean occupation number follows $n_{i,P}=[1-\exp(-2n_i)]/2$ after the parity projection~\cite{Bakr2009,Sherson2010}. 
In contrast, at higher lattice depth ($24E_{\rm r}$), we observe plateaus of constant integer site occupations with almost vanishing fluctuations, directly observing the Mott insulator.

To demonstrate the scattering length tunability of $^7$Li atoms using the Feshbach resonance, we create the Mott insulator under various magnetic fields. 
Even with the same atom number, we observe the central filling $n$ of the Mott insulator varies from $n=3$ to $n=1$ [Fig.~\ref{FigMI}(b)-(d)] as the magnetic field increases.
In particular, with a large scattering length $a_s=680a_B$, we can generate a large size unity filling Mott insulator having 2,500 atoms [Fig.~\ref{FigMI}(d)] with a diameter of 60 sites. 
Such a large system size can reduce finite size effects and could be essential for measuring critical exponents near phase transitions.
Increasing the magnetic field further to have $a_{s}>700 a_B$, we observe significant atom loss and heating so that a cold Mott insulator cannot be created. 
In this regime, the lifetime of the atoms strongly depends on the scattering length and the harmonic curvature, indicating that we are limited by the three-body atom loss~\cite{Pollack2009b}.

We determine the parameters in the Hubbard Hamiltonian, $J$(nearest neighbor tunneling energy) and $U$ (onsite interaction energy) by performing lattice modulation spectroscopy ~\cite{Thilo2004}. 
The tunneling strength can be estimated by measuring the band gap at different lattice depths, where we observe three peaks in the second excited bands. 
Making almost the same intensities of the two incoming lattice beams [lattice1 and lattice2 beam in Fig.~\ref{Fsetup}(a)] and maximizing the beam overlaps of the retro-reflected lattice beams, we achieve nearly isotropic tunneling energy in the square lattice (anisotropy is less than 10$\%$).
Similarly, we determine the on-site interaction energy by directly observing doublon formation when the lattice modulation frequency becomes $\hbar\omega=h\times U$. 

In the best condition, the temperature is estimated to $k_BT = 0.083(1)U$, which is comparable to the spin exchange energy in a 1D system~\cite{Jepsen2020} assuming that the $T/U$ is fixed at different lattice depths. 
We expect that the temperature of the atoms can be reduced by improving the preparation sequence of the Mott insulating phase. 
For example, the condensate size before the lattice ramp up is larger than the size of the Mott insulator (Fig.~\ref{FigMI}) so that the slow thermalization time in the insulating phase might lead to holes~\cite{Hung2010}.
Another direction for reducing temperature is potential engineering that redistributes the entropy of the system~\cite{Bernier2009}, which is reported in Fermi gases of $^6$Li atoms~\cite{Mazurenko2017,Chiu2018}.


\section{\label{Conclusion}Conclusions and outlook}

In conclusion, we have achieved a single-site and single-atom resolved imaging of a strongly interacting bosonic Mott insulator of $^7$Li atoms. 
Employing Raman sideband cooling in a deep square lattice, we collect on average 4,000 photons per atom in a second and have imaging fidelity of 95\% in a large area of 80$\times$80 sites. 
Tuning the interaction near the Feshbach resonance, we are able to produce a cold unity filling Mott insulator with 2,000 atoms at temperature of $T=0.083(1)U/k_B$. 
The large-sized system might allow us to study a finite size scaling, playing a central role in the studies phase transitions. 
Our imaging system enables measurement of the density and spin correlations at arbitrary distances and can be utilized to address individual atoms in each lattice site. 

With this new platform, we could study various phases in a multi-component Bose-Hubbard Hamiltonian~\cite{Kuklov2003,Duan2003,Ehud2003,Kuklov2004}, such as exotic superfluid phases~\cite{Kuklov2003,Kuklov2004} and magnetic ordered states~\cite{Duan2003,Ehud2003}, and the dynamics of spin correlation in the XXZ spin-1/2 Heisenberg Hamiltonian~\cite{Mathy2012,Fukuhara2013b,Ljubotina2017,Gopalakrishnan2019}.
Furthermore, tuning the magnetic field to have an attractive interaction, we could observe stable discrete solitons in higher dimensional optical lattices~\cite{Efremidis2003,Ahufinger2004,Baizakov2004} and  a macroscopic entangled state over different lattice sites~\cite{Jack2005,Buonsante2005}.
Lastly, strong three-body atom loss, which is often avoided, could be utilized to engineer effective hard-core three-body interactions, and intriguing superfluid phases could be observed in the attractive side of the Feshbach resonance~\cite{Daley2009,Diehl2010,Bonnes2011}.


\begin{acknowledgments}
We acknowledge discussion with I. Bloch, C. Gross, J. Koepsell, A. Omran, G. Salomon, P. Schau\ss, and Y. Takahashi. We are funded by Samsung Science and Technology Foundation BA1702-06, National Research Foundation of Korea Grant No.2019M3E4A1080401 and 2020R1C1C1010863. K. Kim is supported by KAIST up program.
\end{acknowledgments}

\bibliography{Ref_Li7QGM}


\end{document}